*Original Article*

# Designing an AI-Powered Mentorship Platform for Professional Development: Opportunities and Challenges

Rahul Bagai[1], Vaishali Mane[2]

*[1]Meta Platforms, Inc.,*
*[2]Google Inc.*

***Abstract** - This article examines the promising prospects and potential hurdles associated with the development of MentorAI, a conceptual AI-driven mentorship platform for professional growth yet to be actualized. The article explores the essential characteristics and technological underpinnings required for the successful creation and efficacy of the MentorAI platform in providing tailored mentorship experiences. The article highlights the transformative potential of MentorAI on various dimensions of professional growth, such as boosting career progression, nurturing skill development, and supporting a balanced work-life environment for professionals. MentorAI, through its AI-based approach, aspires to offer real-time guidance, resources, and assistance customized to each individual's specific needs and goals. Furthermore, the article examines the core technologies crucial to MentorAI's operation, including artificial intelligence, machine learning, and natural language comprehension. These technologies will empower the platform to process user inputs, deliver context-sensitive responses, and dynamically adjust to user preferences and objectives. The deployment of MentorAI presents potential challenges and ethical concerns, as with any groundbreaking technology. The article outlines critical issues like data protection, security, algorithmic bias, and moral quandaries concerning substituting human mentors with AI systems. Addressing these challenges proactively and deliberately is vital to ensure a positive impact on users.*

## 1. Introduction

Mentorship is crucial in professional growth, encompassing a nurturing relationship between an experienced mentor and a less seasoned mentee. Mentorship's significance in professional development can be attributed to several key aspects:

- Skill enhancement: Mentors help mentees recognize their strengths and weaknesses and cultivate vital skills for career progression. Mentors enable mentees to gain valuable expertise in their respective fields through advice and shared knowledge.
- Networking possibilities: Mentors frequently expose mentees to professional networks, forging connections with prominent industry figures—these relationships pave the way for new career prospects, partnerships, and the exchange of ideas.
- Confidence reinforcement: Mentors offer support, constructive criticism, and a secure environment for discussing obstacles, empowering mentees to develop self-assurance, tackle self-doubt, and effectively steer their careers.

- Career counsel: Mentors provide priceless career advancement recommendations, employment opportunities, and professional decision-making wisdom. Their guidance assists mentees in making well-informed choices aligned with their long-term objectives.
- Work-life equilibrium: Mentorship transcends professional development, with mentors advising on achieving a balanced work-life dynamic. By imparting stress management techniques, prioritization, and boundary establishment, mentors enable mentees to flourish both personally and professionally.

Despite the benefits of conventional mentorship schemes, they often require enhancements in their efficacy for promoting professional growth. Common issues include restricted mentor accessibility, time limitations, mismatches, geographical obstacles, insufficient personalization, ineffective scalability, and challenges in evaluating success. These difficulties can obstruct the overall influence of traditional mentorship endeavors.

Alternative mentorship strategies, such as the MentorAI platform, can address these constraints and boost mentorship program efficiency. A MentorAI platform utilizes cutting-edge AI technologies to deliver tailored guidance and backing to professionals, overcoming obstacles traditional mentorship programs face. Employing AI, machine learning, and natural language comprehension, an AI-powered mentorship platform tailors the mentorship experiences to conform to each user's requirements and aims. Accessible round-the-clock and from any location, professionals can obtain real-time counsel and backing via preferred communication channels and at their convenience.

The AI-centric approach enables MentorAI to scale effectively throughout vast organizations, supplying mentorship assistance to numerous professionals without overburdening resources. The platform monitors user progress and generates quantifiable indicators, yielding valuable insights into the mentorship experience's efficacy. By capitalizing on AI technology, MentorAI offers an appealing substitute to conventional mentorship programs, amplifying the overall impact of professional development initiatives.

This article investigates multiple facets of the MentorAI platform and how it can revolutionize professional growth. The article will probe the following domains:

- Examine the potential prospects and obstacles to developing a MentorAI platform.
- Investigate the essential characteristics and fundamental technologies for efficient AI-powered mentorship, including natural language comprehension, machine learning, and user profiling.
- Assess how the MentorAI platform can surmount the drawbacks of traditional mentorship programs, especially in terms of personalization.
- Scrutinize the potential influence of MentorAI on employee satisfaction, involvement, and retention within organizations.
- Suggest a framework for deploying the MentorAI platform in diverse professional contexts, such as large organizations, remote workspaces, and sector-specific domains.
- Appraise the ethical considerations and potential hazards of employing AI technology in mentorship and professional development situations.

By exploring these aspects, this article intends to thoroughly comprehend the MentorAI platform's potential in transforming professional growth and overcoming the limitations of traditional mentorship schemes.

## 2. Key Features and Technologies for AI-Powered Mentorship

The MentorAI platform must incorporate several essential features and components to facilitate personalized, engaging, and impactful mentorship experiences. Some crucial elements and technologies include

- Conversational AI Interface: A MentorAI platform should equip itself with an intuitive conversational AI interface that comprehends and reacts to user inputs in natural language. This interface enables users to interact with the platform as if conversing with a human mentor, fostering a sense of connection and engagement. The conversational AI should be capable of handling complex queries, providing relevant and context-aware responses, and maintaining a consistent and coherent dialogue with the user.
- Personalization Capabilities: The MentorAI platform must include robust personalization capabilities to deliver tailored mentorship experiences to users. By analyzing user inputs, preferences, and interaction history, the platform should be able to adapt its recommendations, guidance, and support to suit each individual's unique needs and goals. This personalization should extend to various aspects of the mentorship experience, such as content recommendations, learning paths, and communication styles.
- AI and Machine Learning: The backbone of any MentorAI platform is the underlying AI and machine learning technologies that drive its functionality. AI algorithms should be capable of learning from user interactions, refining their understanding of user needs and preferences, and continuously improving the platform's performance. Machine learning methodologies, such as supervised and unsupervised, reinforcement, and deep learning, can improve the platform's ability to offer relevant and practical mentorship guidance.
- Natural Language Understanding (NLU): NLU is critical for interpreting and processing user inputs in natural language, allowing the platform to accurately understand and respond to user queries. NLU technologies, such as sentiment analysis, entity recognition, and intent classification, enable the MentorAI platform to comprehend the nuances of human language and generate context-aware, meaningful responses.
- User Profiling and Data Analysis: AI-powered platforms must be able to construct detailed user profiles based on various data points, such as user inputs, preferences, interaction history, and demographic information, to create personalized mentorship experiences. By analyzing this data, the platform can identify patterns, trends, and correlations that inform its mentorship guidance and recommendations, ensuring that the provided support is relevant and effective for each user.
- Integration with External Resources: An effective MentorAI platform should be capable of integrating

with external resources, such as online courses, learning materials, and professional networks. This integration allows the platform to provide comprehensive and up-to-date information, recommendations, and guidance, ensuring users access the most relevant and valuable resources for their professional development.

- Privacy and Security: Ensuring the privacy and security of user data is crucial for the MentorAI platform. Robust encryption, secure data storage, and strict access controls should be in place to protect user information and maintain compliance with data protection regulations. Additionally, the platform should incorporate transparent data usage policies and provide users with control over their data and privacy settings.

By incorporating these key features and technologies, the MentorAI platform can deliver personalized, engaging, and effective mentorship experiences that overcome the limitations of traditional mentorship programs and drive professional development for a diverse range of users.

## 3. Potential Benefits and Impact

The MentorAI platform holds the potential to significantly revolutionize the professional development landscape by offering personalized, scalable, and accessible mentorship experiences. The possible advantages and impacts of such a platform on professionals' career advancement, skill development, and work-life balance are numerous:

- Boosted Career Growth: MentorAI platform can supply tailored guidance and suggestions to help professionals achieve their career objectives effectively. This platform can speed up career advancement and enable more informed decision-making by presenting personalized learning paths, job opportunity recommendations, and career growth strategies.
- Enhanced Skill Development: MentorAI platform can identify skill gaps, recommend appropriate resources, and provide targeted advice to help professionals gain and refine new skills. By utilizing machine learning algorithms, this platform can continually refine its comprehension of user needs and preferences, ensuring that the provided support remains relevant and practical as the user's skill set evolves.
- Improved Work-Life Balance: MentorAI platform can offer advice on stress management, priority-setting, and sustaining a healthy work-life balance. By delivering personalized suggestions and strategies, this platform can assist professionals in navigating their careers' challenges without compromising personal well-being.
- Increased Employee Satisfaction: By offering customized mentorship experiences addressing individual needs and objectives, the MentorAI platform can significantly enhance employee satisfaction. When professionals feel supported in their career development and work-life balance, they are more likely to be content and engaged.

- Greater Employee Engagement: MentorAI platform can nurture a sense of connection and engagement by providing personalized, context-aware support and guidance. By emulating the experience of interacting with a human mentor, this platform can foster a sense of belonging and investment in one's professional growth.
- Higher Employee Retention: Employees who feel supported in their professional development and experience a healthy work-life balance are more likely to remain committed to their organizations. The MentorAI platform can boost employee retention by addressing professionals' individual needs and goals and cultivating a supportive and engaging work environment.

The MentorAI platform can significantly impact professionals' career advancement, skill improvement, and work-life balance by leveraging AI and machine learning capabilities. This platform can enhance employee satisfaction, engagement, and retention, creating a more supportive, nurturing, and productive work environment.

## 4. Barriers to Implementation and Ethical Considerations

Organizations must confront challenges and ethical concerns when implementing a MentorAI platform to ensure smooth integration and a positive user impact. Critical obstacles and ethical considerations include the following:

- Data Privacy and Security: MentorAI platform depends on vast user data for personalized guidance and support. Guaranteeing the privacy and security of this sensitive data is vital for maintaining user trust and adhering to relevant data protection regulations. Organizations must implement robust data privacy and security protocols, including data encryption, access control, and secure storage and transmission methods, to safeguard user information.
- Algorithmic Bias: AI algorithms and machine learning models can be prone to bias from imbalanced training data or flawed algorithms. Algorithmic bias can result in unfair or discriminatory outcomes for specific user groups, which compromises the effectiveness of the mentorship platform and may harm users. Organizations should invest in transparent and accountable AI development processes to tackle this issue, ensuring their algorithms and models are thoroughly evaluated for fairness, precision, and inclusiveness.
- Ethical Use of AI: As the MentorAI platform gains popularity, concerns may arise regarding the potential for AI to replace human mentors, leading to job displacement and a loss of interpersonal connections in the mentoring process. Developing and implementing the MentorAI platform ethically is crucial, emphasizing that the technology should supplement and improve human mentorship rather than replace it. Organizations should encourage a balanced approach, integrating AI-driven guidance and support alongside traditional mentorship, to maximize the advantages of both methods.

- Transparency and Explainability: Users may have concerns about the clarity and explainability of AI-driven recommendations and guidance, particularly if they feel the platform's decisions should be more precise and understandable. Organizations should prioritize developing transparent and explainable AI systems, ensuring that users comprehend the rationale behind the platform's suggestions and can make informed decisions about their professional development.

- Emotional Intelligence and Empathy: MentorAI platform may need help replicating human mentors' emotional intelligence and empathy to the mentorship relationship. This limitation could affect the platform's ability to effectively support users through personal and emotional challenges related to their professional development. Organizations should invest in advanced natural language understanding and sentiment analysis technologies to address this issue, allowing the AI system to better identify and react to users' emotional needs and provide more compassionate guidance.

By recognizing and addressing these obstacles and ethical considerations, organizations can develop and implement a MentorAI platform that improves the overall efficacy of professional development initiatives while maintaining user trust and adhering to ethical standards.

# 5. Proposed Design and Implementation Framework

The design and implementation of a MentorAI platform can be structured using the Define, Design, Develop, Deliver, and Monitor & Maintain framework to ensure its effectiveness and seamless integration into organizations. The proposed framework consists of the following stages:

### 5.1. Define
*5.1.1. Define Objectives and Scope*

Clearly outline the objectives and scope of the MentorAI platform, including the target audience, desired outcomes, and key performance indicators. Engage relevant stakeholders, such as HR teams, organizational leaders, and potential users, in defining the project's goals and expectations.

### 5.2. Design
*5.2.1. Select and Develop Key Technologies*

Identify and develop the core AI, machine learning, and natural language understanding technologies underpinning the mentorship platform. This stage involves selecting appropriate algorithms, designing and training models, and developing a conversational AI interface to facilitate user interaction.

*5.2.2. Design Personalization Mechanisms*

Develop mechanisms for personalizing the mentorship experience, such as user profiling, adaptive learning paths, and context-aware recommendations. These features should cater to user needs, preferences, and goals, ensuring the platform remains relevant and engaging.

*5.2.3. Design User Interface and Experience*

Design an intuitive and user-friendly interface for the mentorship platform, ensuring users can easily access and navigate the platform's features and content. This stage should also involve usability testing and iterative design improvements to enhance the user experience.

### 5.3. Develop
*5.3.1. Integrate Relevant Content and Resources*

Source and integrate a wide range of content and resources to support various aspects of professional development, such as skill-building materials, career guidance articles, and work-life balance tools. These resources should be curated and organized to facilitate easy access and navigation by users.

*5.3.2. Implement Data Privacy and Security Measures*

Develop strong data privacy and security protocols to safeguard user data and adhere to applicable regulations. This stage involves data encryption, access controls, and secure storage and transmission methods to protect sensitive user data.

### 5.4. Deliver
*5.4.1. Pilot and Refine the Platform*

Launch a pilot version of the MentorAI platform, allowing users to test its features and provide feedback. This feedback will refine the platform's design, features, and technologies, meeting user needs and expectations.

### 5.5. Monitor and Maintain
*5.5.1. Monitor and Evaluate Outcomes*

Continuously evaluate the platform's performance using key performance indicators, user feedback, and analytics data. This stage involves identifying areas for improvement, optimizing the platform's features and algorithms, and measuring its impact on users' professional development and organizational outcomes.

*5.5.2. Update and Maintain the Platform*

Regularly update its content, resources, and features to stay current and relevant. Perform maintenance tasks, such as fixing bugs, addressing security vulnerabilities, and updating underlying technologies, to ensure the platform's ongoing stability and performance.

By following this framework, organizations can develop and implement a MentorAI platform that effectively supports professionals in their development while addressing potential challenges and barriers.

To effectively integrate the MentorAI platform into organizations, the following strategies may help overcome implementation barriers:

- Secure Buy-In from Key Stakeholders: Engage organizational leaders, HR teams, and potential users early in the development process to ensure the incorporation of their input and support into the platform's design and implementation.
- Develop a Clear Integration Plan: Establish a clear plan for integrating the mentorship platform into the organization's professional development programs, including timelines, resource allocation, and communication strategies.
- Provide Comprehensive Training and Support: Offer training and support to users to help them understand the platform's features, benefits, and usage guidelines. This may involve developing user guides, hosting workshops, and providing ongoing technical support.
- Promote a Culture of Continuous Learning and Innovation: Promote a culture of ongoing learning and innovation within the organization, highlighting the importance of mentorship and professional growth. This may involve recognizing and rewarding employees who actively engage with the platform and demonstrate progress in their career growth.

By following this proposed design and implementation framework, organizations can successfully develop and integrate the MentorAI platform, enhancing the effectiveness of professional development initiatives and overcoming the limitations of traditional mentorship programs.

# 6. Success Metrics

We can categorize success metrics for a MentorAI platform into quantitative and qualitative measures. These metrics help evaluate the platform's effectiveness, impact on users, and alignment with the desired outcomes. Consider these key success metrics:

*6.1. Quantitative Metrics*
- User engagement: Track the number of active users, session duration, and frequency of use. Higher concentration indicates that users find the platform valuable and relevant.
- Goal achievement: Measure the percentage of users who achieve their stated professional development goals or milestones within a given time frame.
- Skill improvement: Assess the improvement in users' skills or competencies before and after using the platform, using tests, self-assessments, or performance evaluations.
- User growth and retention: Monitor the rate at which new users join the platform and the retention rate of existing users over time.
- Network growth: Track the expansion of users' professional networks, including new connections, collaborations, and introductions made through the platform.
- Career progression: Measure the rate of promotions, role changes, or other career advancements among users who have engaged with the platform.
- Platform satisfaction: Gather user feedback through surveys or ratings to evaluate their overall satisfaction with the platform's features, content, and user experience.

*6.2. Qualitative Metrics*
- User testimonials: Collect user testimonials and case studies that showcase the platform's positive impact on their professional development, career growth, and work-life balance.
- Mentor-mentee relationship quality: Assess the quality of the relationships formed between the AI mentor and users, including factors such as trust, communication, and rapport.
- Personal growth: Evaluate the platform's impact on users' development, including increased self-confidence, resilience, and self-awareness.
- Organizational benefits: Gather feedback from organizations implementing the platform to understand its impact on employee satisfaction, engagement, and retention.
- Ethical considerations: Assess the platform's adherence to ethical principles, such as data privacy, security, and addressing algorithmic biases.

Continuously tracking and assessing these success metrics offers crucial insights into the MentorAI platform's effectiveness, guiding ongoing enhancement efforts and guaranteeing that the platform achieves the intended results for users and organizations.

## 7. Use Cases

MentorAI can be used in various settings and industries to support professionals in their growth and development. Some possible use cases include:

- Educational Institutions: Universities and colleges can integrate MentorAI into their career services offerings, helping students and alums identify career paths, develop relevant skills, and access networking opportunities.
- Corporate Organizations: Companies can use MentorAI to supplement their internal mentorship programs, providing personalized guidance to employees for skill development, performance improvement, and leadership growth.
- Professional Associations: Industry-specific associations can offer MentorAI as a membership benefit, helping professionals in their field navigate career challenges, develop industry-specific skills, and expand their professional networks.
- Government Agencies: Government agencies can employ MentorAI to support public sector employees in their professional development, ensuring they remain up-to-date with industry trends, regulations, and best practices.
- Nonprofit Organizations: Nonprofits can leverage MentorAI to help their staff and volunteers develop the skills and knowledge needed to serve their communities and advance their mission effectively.
- Online Learning Platform: MentorAI can be integrated with an online learning platform, providing personalized mentorship and guidance to users as they engage with course content, enhancing the overall learning experience.
- Career Coaching Services: Professional career coaches can utilize MentorAI to supplement their services, offering clients additional personalized guidance and resources to support their career growth.
- Remote Work Environments: Companies with remote or hybrid workforces can use MentorAI to support their employees as they navigate the unique challenges of remote work, including communication, collaboration, and work-life balance.

By integrating MentorAI into various environments, organizations and individuals can harness the power of personalized guidance and support to enhance professional development and growth.

## 8. Conclusion

In conclusion, MentorAI offers promising opportunities to revolutionize professional development by overcoming traditional mentorship limitations. Utilizing advanced AI technologies, machine learning, and natural language understanding, this platform delivers personalized, accessible, and scalable mentorship experiences for diverse professionals. Benefits include enhanced career growth, skill development, work-life balance, and improved employee satisfaction, engagement, and retention.

However, design and implementation challenges, such as data privacy, security, algorithmic bias, and ethical considerations, must be addressed responsibly to ensure AI-powered solutions complement, not supplant, human mentorship. Continued research and development can unlock new professional development possibilities, empowering individuals and organizations to harness their potential and create a more inclusive, supportive landscape for professional growth.